\documentclass[prd,nofootinbib,preprint,superscriptaddress,a4paper]{revtex4-1}

\pdfoutput=1

\usepackage{graphicx}
\usepackage{hyperref}
\usepackage{epsfig}
\usepackage{amsmath}
\usepackage{amsthm}
\usepackage{amssymb}
\usepackage{color}

\newcommand{\be}{\begin{equation}}
\newcommand{\ee}{\end{equation}}

\usepackage[utf8]{inputenc}

\renewcommand\[{\begin{equation}}
\renewcommand\]{\end{equation}}

\newcommand{\ba}{\begin{eqnarray}}
\newcommand{\ea}{\end{eqnarray}}

\newcommand{\Kc}{{\cal K}}

\newcommand{\Fc}{{\cal F}}
\newcommand{\Lc}{{\cal L}}

\begin{document}
        
\title{Convergent sum of EFT corrections to Schwarzschild metric requires UV locality}

\author{Yang Liu}
\email{liuyang2023@shanghaitech.edu.cn}
\affiliation{
School of Physical Science and Technology, ShanghaiTech University, 201210 Shanghai, China
}

\author{Alexey S. Koshelev}
\email{askoshelev@shanghaitech.edu.cn}
\affiliation{
School of Physical Science and Technology, ShanghaiTech University, 201210 Shanghai, China
}
\affiliation{
	Departamento de F\'isica, Centro de Matem\'atica e Aplica\c{c}oes (CMA-UBI),
	Universidade da Beira Interior, 6200 Covilh\~a, Portugal }

\author{Anna Tokareva}
\email{tokareva@ucas.ac.cn}
\affiliation{School of Fundamental Physics and Mathematical Sciences, Hangzhou Institute for Advanced Study, UCAS, Hangzhou 310024, China}
\affiliation{International Centre for Theoretical Physics Asia-Pacific, Beijing/Hangzhou, China}
\affiliation{Theoretical Physics, Blackett Laboratory, Imperial College London, SW7 2AZ London, U.K.}

\author{Ziyue Zhu}
\email{zhuziyue25@mails.ucas.ac.cn}
\affiliation{School of Fundamental Physics and Mathematical Sciences, Hangzhou Institute for Advanced Study, UCAS, Hangzhou 310024, China}

\begin{abstract}
Corrections to vacuum black hole solutions of general relativity (GR) are considered in an effective field theory (EFT) framework, perturbatively in EFT coefficients, focusing on the Schwarzschild solution of GR. We find dominant corrections to the Schwarzschild metric in all orders in the derivative expansion far away from the horizon. These corrections can be summed up in a closed form through EFT coefficients up to all orders in derivatives and to the second order in curvature. It occurs that such a summation is convergent only for localizable theories, making a direct connection between the graviton scattering amplitudes properties and the applicability of a perturbative treatment of an EFT of gravity. We further apply our results to logarithmic form-factors which appear in the 1-loop effective action for GR in four dimensions. We find out that the corresponding corrections to the Schwarzschild metric are stronger than those from the tree-level EFT operators. The developed framework can be extended to account for the corrections to the other BH solutions in GR, such as the Kerr metric.
\end{abstract}

\maketitle

\section{Introduction}

Corrections to General Relativity (GR) are inevitable and appear in many contexts. Construction of the 1-loop effective action  \cite{Deser:1976yx,Fradkin:1978yw,Fradkin:1978yf,Barvinsky:1983vpp} to GR revealed the appearance of higher curvature and higher derivative, and even non-local operators. String corrections suggest that higher curvature corrections naturally arise \cite{Zwiebach:1985uq}. An Effective Field Theory (EFT) approach \cite{Weinberg:1980wa} to GR \cite{Donoghue:1994dn} generates an infinite tower of operators of all positive mass dimensions arranged in Hilbert series \cite{Ruhdorfer:2019qmk}. The more recent idea of asymptotic safety in gravity \cite{Dou:1997fg,Percacci:2007sz,Reuter:1996cp,Reuter:2001ag,Knorr:2019atm} brings out a construction based on the RG flow considerations and the existence of an asymptotically safe fixed point. This construction consists of various terms in addition to an initial GR action arranged with the help of operator functions of the d'Alembertian, the so-called form-factors. Moreover, an idea of higher and infinite derivative gravity theories was developed by many authors in an attempt to pinpoint a unitary and renormalizable gravity theory \cite{Stelle:1976gc,Krasnikov:1987yj,Krasnikov:2024jld,Kuzmin:1989sp,Tomboulis:1997gg,Modesto:2011kw,Biswas:2016egy}.

Presence of new terms in an action nearly always implies that known solutions should be modified. In particular, known black hole (BH) solutions of GR are expected to change upon inclusion of GR corrections.
Despite several known results (see \cite{Boulware:1985wk} for a BH solution with a Gauss-Bonnet term in 5 dimensions and a short review \cite{Garraffo:2008hu} regarding exact BH solutions in Lanczos-Lovelock models of gravity \cite{Padmanabhan:2013xyr}), solving gravitational field equations in generalized theories of gravity is known to be a problem of immense difficulty. Very interesting results were obtained in quadratic gravity \cite{Stelle:1976gc} combining analytic and perturbative methods \cite{Lu:2015cqa,Lu:2015psa}.
However, in a case when corrections to GR are treated perturbatively, one can also implement perturbative schemes to find corrections to known GR solutions. In this case coefficients of the GR corrections play the role of small perturbation parameters. Perturbative corrections to BH solutions in such an approach in an EFT of GR were intensively studied. Remarkable results concerning the corrections to the metric of vacuum BH solutions are presented in \cite{ Kats:2006xp,Cheung:2016yqr,Melville:2024zjq,Arkani-Hamed:2021ajd,Bueno:2024dgm,Bueno:2025qjk,Wang:2022sbp,Boyce:2025fpr,Miguel:2023rzp}. Recently, such corrections for the case of near extremal Kerr metric received a lot of attention due to the significant modifications in the observables \cite{Horowitz:2023xyl}.
Finding and analyzing the dominant perturbative corrections to all orders in an EFT of gravity expansion far away from the horizon to BH solutions in GR using Schwarzschild as the primary departure point comprises the context of the present paper.

It is clear that in the case of EFT expansion, a set of operators determined by the Hilbert series represents a derivative expansion. It means that the perturbations are, in fact, higher derivative terms compared to those already present in the model. This can give more solutions to equations of motion, including unstable ones. In the EFT context, one should be careful not to take such solutions into account, as they usually cannot be properly described by EFT, since the EFT description will break. The very existence of such unstable solutions depends on the UV completion, which is unknown in the case of gravity. Moreover,  even for small corrections to GR solutions, it is not guaranteed that the perturbative construction of the solution is a convergent procedure. There are two reasons for this.

The first reason is as follows. Consider the EFT corrected action truncated to a certain order of mass dimension and derivatives. As the equations are nonlinear, the next order in EFT couplings of the truncated action may be larger than the previous one for certain values of the spacetime coordinates (presumably, in the near-horizon limit of a BH). This can well be the case of near-extremal BH-s, where linear order in EFT couplings naively predicts a near-horizon EFT breakdown \cite{Horowitz:2024dch}. However, it seems to be more accurate to expect that in reality, this behavior corresponds to the breaking of the linear approximation of the perturbative scheme of constructing a corrected solution \cite{Horowitz:2022mly,Horowitz:2023xyl,Horowitz:2024dch,Horowitz:2024kcx,Chen:2024sgx,DelPorro:2025fiu}.

The other reason is connected to the truncation of the infinite sequence of the EFT operators. The question is whether it is correct to assume that only a finite subset of operators is responsible for the dynamics at low energies, even below the cutoff scale. It is important to note at this point that consistent EFTs having unitary, causal, and local UV completion must have an infinite subset of operators leading to the $s^n$ terms in the forward limit ($t\to0_-$) of the $2\rightarrow 2$ scattering amplitudes \cite{Adams:2006sv} ($s,t,u$ are standard Mandelstam variables here). These terms are constrained by positivity bounds and are required to be present without a large hierarchy among them \cite{Hong:2023zgm}. In general, corrections to BH solutions generated by EFT operators should form a power series in $1/r$. One would naively expect that for a large enough BH, only the first corrections are important. But the situation can be more subtle because under a certain choice of the sequence of EFT couplings, these series can have too small (not allowing to continue the solution from spatial infinity to the horizon) or even zero radius of convergence.

In our work, we focus on studying whether and when a perturbative approximation scheme for finding a vacuum BH solution in the EFT of gravity will be well defined and result in a physically meaningful outcome. We mainly consider an EFT parametrization capturing, among any given mass dimension, all the operators with the maximal number of derivatives acting on a metric field. Such operators are naturally expected to provide dominant corrections to BH solutions among those of the same mass dimension, and these operators have the form
\begin{equation}
    {\cal O}_n=C_{\mu\nu\lambda\rho}\,\frac{\square^n}{\Lambda^{2n}}\,C^{\mu\nu\lambda\rho}
    \label{kind1}
\end{equation}
for $n\geq0$ where $C_{\alpha\beta\gamma\delta}$ is the Weyl tensor, $\Box$ is the covariant d'Alembert operator, and $\Lambda$ is the mass parameter associated with the EFT cut-off scale.
Although by a sequence of transformations (including the use of the Bianchi identity and field redefinitions), it is possible to rewrite it in a familiar form like in EFT of GR \cite{Ruhdorfer:2019qmk} (where it is also explained why only the powers of the Weyl tensor ad covariant derivatives are relevant), we will be using the just written form in order to capture the dominant corrections.
Such operators allow us to make a link to the properties of the non-perturbative graviton propagator, as they do change the perturbative propagator, while operators with higher degree in the Weyl tensor obviously do not \cite{Biswas:2016etb}. Remarkably, we will find a tight relation between the very possibility of constructing a BH solution perturbatively and the property of UV locality of the graviton propagator and amplitudes.
Additionally, it is possible to consider a series of such corrections
\begin{equation}
    \sum_{n\geq0}{f_C}_nC_{\mu\nu\lambda\rho}\,\frac{\square^n}{\Lambda^{2n}}\,C^{\mu\nu\lambda\rho}=C_{\mu\nu\lambda\rho}\,\Fc({\square}/{\Lambda^2})\,C^{\mu\nu\lambda\rho}
    \label{kind1sum}
\end{equation}
and perform a resummation of the contributions of such operators for a BH solution. Here ${f_C}_n$ are dimensionless coefficients which are EFT coefficients and are to be treated perturbatively.

On the other hand, a similar kind of operators comprises corrections arising in the 1-loop effective action for GR. The latter situation is more subtle as it requires an inclusion of an operator with a logarithmic form-factor in four dimensions,
\begin{equation}
    {\cal O}_\text{log}=\gamma(\mu) C_{\mu\nu\lambda\rho}\,\log(\square/\mu^2)\,C^{\mu\nu\lambda\rho}
    \label{kindlog}
\end{equation}
where $\mu$ is the renormalization scale. Being a non-analytic function at the origin, it requires a special treatment.
To start with, even deriving equations of motion is already a problematic task as the latter relies on the Taylor expansion of a form-factor and a convergence of the corresponding series.
We will find that meaningful results can be obtained by analytically continuing results for monomial operators containing $\Box^n/\Lambda^{2n}$, and will show that the 1-loop contribution with a logarithmic form-factor yields the dominant correction to BH solutions compared to those resulting from power law form-factors.

The paper is organized as follows. In Section~II, we set up the model and present the equations of motion to be used in the analysis. In Section~III, we study dominant corrections to Schwarschild solution coming from operators of kinds (\ref{kind1}) and (\ref{kind1sum}). In Section~IV, we study the convergence of various series obtained in Section~III and the physical implications of different sets of EFT parameters. In Section~V, we attack the situation of operators with a logarithmic form-factor (\ref{kindlog}) arising in the 1-loop effective action in GR. We then summarize our findings in the Conclusion Section and outline questions for future study.

\section{Model set-up}

The model under consideration is
\begin{equation}
    S=\int d^Dx\sqrt{-g}\left(\frac{M_P^{D-2}}2R+\frac{\lambda M_P^{D-4}}2 C_{\mu\nu\alpha\beta}\Fc_C(\Box/\Lambda^2)C^{\mu\nu\alpha\beta}\right)
    \label{ourmodel}
\end{equation}
Here $M_P$ is the Planck mass. $\Fc_{C}(\Box/\Lambda^2)$ is an operator function of the d'Alembert operator, a form-factor. It is assumed to be compatible with the IR limit of GR. This does not necessarily imply to be analytic in the origin as for example $\log$-type form-factor does also give a correct IR limit.\footnote{There is a recent analysis with fraction type form-factors also targeting a similar task of constructing perturbatively corrected solutions for BH-s \cite{Borissova:2025nvj}.} Here $\lambda$ is a dimensionless coupling which will be used as the perturbation parameter and thus will play an essential role in our analysis.
$\Lambda$ is the scale of higher derivative modifications, which can have different interpretations as explained in the Introduction.
Note that the perturbative unitrity demands the presense of another term $\frac{\lambda M_P^{D-4}}2 R\Fc_R(\Box/\Lambda^2)R$ and moreover the form-factors should be related as $\Fc_R(\Box) +\frac{D-3}{D-1}\Fc_C(\Box)=0$ (more details in case of $D=4$ can be found in \cite{Koshelev:2025pxg}). However, this extra term is irrelevant to our analysis, as it will be demonstrated in the next paragraph.

Our goal is to find perturbatively in coupling $\lambda$ corrections to existing BH solutions of GR.
Terms quadratic and of a higher order in $R$ (or $R_{\mu\nu}$) can be neglected as they will vanish in the perturbative scheme in the linear order in $\lambda$. In this paper, we will concentrate on the linear order in perturbations and, therefore, can neglect such operators. Moreover, we will build the analysis looking for corrections to the Schwarzschild metric, leaving other important solutions, including the Kerr metric for a rotating BH, for future projects.
One naturally expects the new solutions to be BH-s, but we should exercise some caution since we have to demonstrate that the perturbative approach provides solutions that can be extended to the horizon. It is clear that the suggested approach is in line with the paradigm of an EFT and is very much justified physically, since we are discussing corrections to IR results considering UV model corrections.

Equations of motion for action (\ref{ourmodel}) in the vacuum have the form
\begin{equation}
\begin{split}
E^\mu_\nu&=-M_P^{D-2}G^\mu_\nu+\lambda M_P^{D-4}\Delta^\mu_{\nu}\\
&=-M_P^{D-2}G^\mu_\nu\\
&+\lambda M_P^{D-4}\left[
\frac12\delta^\mu_\nu C_{\gamma\delta\alpha\beta}
\Fc_C(\Box/\Lambda^2)C^{\gamma\delta\alpha\beta}-2C^\mu_{\delta\alpha\beta}\Fc_C(\Box/\Lambda^2)C_\nu^{\phantom{\nu}\delta\alpha\beta}\right.\\
&+2\left(\frac{2}{D-2}R_{\alpha\beta}
+2\nabla_\alpha\nabla_\beta\right)
\Fc_C(\Box/\Lambda^2)C_{\nu}^{\phantom{\nu}\alpha\beta\mu}\\
&+\left.{\Lc_C}^\mu_\nu
-\frac{1}{2}\delta^\mu_{\nu}\left({\Lc_C}^\sigma_\sigma+\bar\Lc_{C}
\right)+4{\Kc_C}^\mu_\nu\right]=0
\end{split}
\label{ourEOM}
\end{equation}
where
\begin{equation*}
{\Lc_C}^\mu_\nu=\sum_{n=1}
^\infty
\frac{{f_C}_n}{\Lambda^{2n}}\sum_{l=0}^{n-1}(\nabla^\mu {\Box^l{C}
}{}^\alpha_{\phantom{\alpha}\beta\gamma\delta}) 
(\nabla_\nu  {\Box^{n-l-1}{C}}{}_\alpha^{\phantom{\alpha}\beta\gamma\delta})
,~\bar\Lc_C=\sum_{n=1}
^\infty
\frac{{f_C}_n}{\Lambda^{2n}}\sum_{l=0}^{n-1}({\Box^l C}{}^\alpha_{\beta\gamma\delta})
({\Box^{n-l}C}{}_\alpha^{\phantom{\alpha}\beta\gamma\delta}),
\end{equation*}
\begin{equation*}
{\Kc_C}^\mu_{\nu}=
\sum_{n=1}
^\infty
\frac{{f_C}_n}{\Lambda^{2n}}\sum_{l=0}^{n-1}\nabla_\beta[
({\Box^l{C}}{}^{\beta\rho}_{\phantom{\beta\rho}\gamma\zeta})
(\nabla^\mu{\Box^{n-l-1}{C}}{}^{\phantom{\nu\rho}\gamma\zeta}_{\nu\rho})
-(\nabla^\mu
{\Box^l{C}}{}^{\beta\rho}_{\phantom{\beta\rho}\gamma\zeta})
({\Box^{n-l-1}{C}}{}^{\phantom{\nu\rho}\gamma\zeta}_{\nu\rho})]
\end{equation*}
and everywhere a symmetrization with respect to free indices is assumed. The quantity $\Delta^\mu_\nu$ is self-explanatory.
Note that the above equations are obtained under an essential assumption that the form-factor is Taylor expandable at the origin and coefficients ${f_C}_n$ are introduced as follows
\begin{equation}
\Fc_{C}(\Box/\Lambda^2)=\sum\limits_{n\geq0}{f_{C}}_n\Box^n/\Lambda^{2n}
\label{taylorF}
\end{equation}
While this is a sufficient condition to restore the IR limit of GR (not necessary though), one may want to consider a wider class of functions. The most notable and already mentioned is $\Fc(\Box/\Lambda^2)\sim\log(\Box/\Lambda^2)$ appearing in the 1-loop correction to GR. To accommodate such form-factors, one can write them using some integral representation or limits which allows to do the Taylor expansion first and perform the corresponding representation integration later. We will specifically discuss the case of the logarithm in Section~\ref{seclog}.

Moreover, we have neglected potentially infinitely many boundary terms. We can do this in our analysis as long as we are interested in modifications to GR solutions far away from an event horizon. Such terms may become important, however, for consideration of the thermodynamics of the obtained solutions --- an important topic which we defer to a future study.

\section{Corrections to Schwarzschild solution far away from the horizon}

The Schwarzschild metric in $D$ dimensions can be written in Schwarzschild coordinates as
\begin{align}
    ds_0^2 =  - \left( 1 - \left(\frac{r_D}{r}\right)^{D-3} \right)d{t^2} + \left( 1 - \left(\frac{r_D}{r}\right)^{D-3} \right)^{-1}d{r^2} + {r^2}d\Omega^2_{D-2} 
    \label{SchD}
\end{align}
where $r_D$ is the Schwarzschild BH radius in $D$ dimensions for which $r_D^{D-3}M_P^{D-2}\sim M$ where $M$ is the BH mass \cite{Zwiebach:2004tj,Robinson:2006yd}. Now we are going to construct corrections to this metric to find a solution to equations (\ref{ourEOM}) perturbatively in $\lambda$. The following parametrization  will be used
\begin{equation}
\begin{split}
    ds^2_\lambda &=  - \left( 1 - \left(\frac{r_D}{r}\right)^{D-3}+\lambda\frac{P_D(r)}{r^{D-3}} \right)(1+\lambda F_D(r))d{t^2} \\
    &+ \left( 1 - \left(\frac{r_D}{r}\right)^{D-3}+\lambda\frac{P_D(r)}{r^{D-3}} \right)^{-1}d{r^2} + {r^2}d\Omega^2_{D-2} 
    \end{split}
    \label{SchDlambda}
\end{equation}
Substituting the latter metric in (\ref{ourEOM}), one obviously restores the Schwarzschild solution in GR in the order $\lambda^0$. In the order $\lambda^1$, we have to solve the following equation
\begin{equation}
    -M_P^2\left.\frac{d(\left.G^\mu_\nu\right|_{ds^2_\lambda})}{d\lambda}\right|_{\lambda=0}+ \left.\Delta^\mu_{\nu}\right|_{ds_0^2}=0
    \label{EOMlambda1}
\end{equation}
Upon computing the first term in the latter equation one sees the convenience of the specific parametrization (\ref{SchDlambda}).
The term $G^0_0|_{ds^2_\lambda}$ is always exactly only linear in $\lambda$ and is moreover monomial in $P'_D(r)$ only. Hereafter, prime is a derivative with respect to the radius $r$ and detailed full formulae can be found in Appendix~\ref{appA}.

The $({}^0_0)$ component of equation (\ref{EOMlambda1}) reads
\begin{equation}
M_P^2\frac{D-2}2\frac{{P_D}'(r)}{r^{D-2}}=\frac{r_D^{2(D-3)}}{r^{2D-2}}\sum_{n=0}^\infty\frac{c_n(D) {f_C}_n}{r^{2n}\Lambda^{2n}}\left(1+O\left(\frac{r_D}r\right)\right)
=\frac{r_D^{2(D-3)}}{r^{2D-2}}\Phi(D,r)\left(1+O\left(\frac{r_D}r\right)\right)
\label{EOMlambda100}
\end{equation}
Here we use one more approximation, assuming that we are far away from the horizon and $\Phi(D,r)$ is a short-hand for the sum in the RHS of the equation.
 We recall that coefficients ${f_C}_n$ are the theory parameters as they determine the form-factor $\Fc_C(\Box/\Lambda^2)$ through expansion (\ref{taylorF}). However, $c_n$ are numeric coefficients coming out as a result of the evaluation of the RHS of equation (\ref{EOMlambda1}) on the $D$-dimensional Schwarzschild metric. As a consequence, they depend on the dimension $D$.
Note that the zero term in the sum on the RHS of equation (\ref{EOMlambda100}) does not have $O(r_D/r)$ corrections.
 Explicit form of coefficients $c_n$ is presented in Appendix~\ref{appA}, equations (\ref{cnD}), (\ref{alphaD}).
 Here we write down the most essential coefficients for clarity. Namely
\begin{equation}
    c_{n\geq2}(D)=2\alpha(D)4^{n}\Gamma(n+\frac{D+1}2) \Gamma(n+D-1)
 \end{equation}
where in four dimensions $\alpha(4)=2/\sqrt{\pi}$ and also note that $c_0(4)=0$, $c_1(4)=90$ and $c_0(D>4)<0$.

Finding out the function $P_D(r)$ far away from the horizon implies a straightforward integration of the series on the RHS of (\ref{EOMlambda100}). It is, however, possible to sum up this series and express it through the form-factor $\Fc_C(\Box/\Lambda^2)$.
To do so we first split $\Phi(D,r)$ as follows
\begin{equation}
\Phi(D,r)=c_0(D) {f_C}_0+\frac{c_1(D) {f_C}_1}{r^{2}\Lambda^{2}}+\sum_{n=2}^\infty\frac{c_n(D) {f_C}_n}{r^{2n}\Lambda^{2n}}=c_0(D) {f_C}_0+\frac{c_1(D) {f_C}_1}{r^{2}\Lambda^{2}}+\Phi_2(D,r)
    \label{sumP}
\end{equation}
This splitting is coming naturally from the fact that coefficients $c_n$ starting from $n=2$ obey a closed form matching through Gamma-functions. Introducing $z={(r\Lambda)^2}/{4}$ one can rewrite $\Phi_2(D,r)$ as
\begin{equation}
    \Phi_{2}\left(D, z\right) =\sum ^{\infty }_{n=2}2\alpha \left( D \right) \Gamma \left( n+\dfrac{D +1}{2}\right) \Gamma \left( n+D -1\right) {f_C}_n z^{-n}
    \label{totransform}
\end{equation}
The key observation at this stage is that Gamma-factors in the latter series can be written using an integral involving the modified Bessel function $K_\nu(x)$ (\ref{intgg})
\begin{equation}
  \Phi_{2}\left(D, z\right)=  4\alpha \left( D \right) z^{\frac{3 D-1}{4}}\sum ^{\infty }_{n=2}\int ^{\infty }_{0}d\tau{f_C}_n\tau^{n+\frac{3D -5}{4}}K_{\frac{D-3}{2}}\left( 2\sqrt{z\tau}\right)
\end{equation}
Note that for even $D$ this is clearly some form of the Laplace transform since the index of the $K$-function is a half-integer and the corresponding modified Bessel function can be re-expressed through powers and an exponent of its argument. Now, one immediately notices that the summation over $n$ can be written as
\begin{equation}
    \sum_{n=2}^{\infty}{f_C}_n\tau^n=\Fc_C(\tau)-{f_C}_0-{f_C}_1 \tau
\end{equation}
Combining all together we have transformed  equation (\ref{EOMlambda100}) to the following form
\begin{equation}
\label{Pprime}
    P'_D\left( r \right) =\dfrac{2}{D -2}\dfrac{1}{M_P^{2}}\dfrac{r_{D}^{ 2(D-3) }}{r^{D}}\cdot \left( c_{0}\left( D\right) {f_C}_{0}+\dfrac{c_{1}\left( D \right) {f_C}_{1}}{r^{2}\Lambda ^{2}}+\Phi _{2}\left( D,r\right) \right)
\end{equation}
where $\Phi_2(D,r)$ has the following integral representation through $\Fc_C(\tau)$
\begin{equation}
  \Phi_{2}\left( D,r\right)=  4\alpha \left( D \right) \left(\frac{r\Lambda}2\right)^{\frac{3 D-1}{2}}\int ^{\infty }_{0}d\tau\left(\Fc_C(\tau)-{f_C}_0-{f_C}_1 \tau\right)\tau^{\frac{3D -5}{4}}K_{\frac{D-3}{2}}\left( r\Lambda\sqrt{\tau}\right)
  \label{phi2int}
\end{equation}
where we have returned back to $r$. Finally, integrating over $r$ one yields
\begin{equation}
    P_D\left( r \right) =-\dfrac{2}{D -2}\dfrac{1}{M_P^{2}}\dfrac{r_{D}^{ 2(D-3) }}{r^{D}}\cdot \left( r\frac{c_{0}\left( D\right) {f_C}_{0}}{D-1}+\dfrac{c_{1}\left( D \right) {f_C}_{1}}{r\Lambda ^{2}(D+1)}+\tilde \Phi _{2}\left( D,r\right) \right)
    \label{Panswer}
\end{equation}
where $\tilde\Phi_2(D,r)$ comes as the result of an integration of $\Phi_2(D,r)/r^D$ with respect to $r$ with subsequent factoring of $-1/r^D$ out. The corresponding expression is
\begin{equation}
\begin{split}
  \tilde\Phi_{2}\left( D,r\right)&=  \frac{4\alpha \left( D \right) }\Lambda\left(\frac{r\Lambda}2\right)^{\frac{3 D-1}{2}}\int ^{\infty }_{0}d\tau\left(\Fc_C(\tau)-{f_C}_0-{f_C}_1 \tau\right)\tau^{\frac{3D -7}{4}}
  K_{\frac{D-1}{2}}\left( r\Lambda\sqrt{\tau}\right)
  \end{split}
  \label{phitilde2int}
\end{equation}
An integration constant must vanish in order not to modify the Newtonian potential at infinity. The reason why this integration went smoothly is that we explicitly hit the property of Bessel functions (\ref{intsimple}).

The $({}^1_1)$ component of equation (\ref{EOMlambda1}) reads
\begin{equation}
\begin{split}
M_P^2
\frac{D-2}2\left[\frac{{F_D}'(r)}{r}\left(1+O\left(\frac{r_D}r\right)\right)+\frac{{P_D}'(r)}{r^{D-2}}\right]
&=\frac{r_D^{2(D-3)}}{r^{2D-2}}\sum_{n=0}^\infty\frac{d_n(D) {f_C}_n}{r^{2n}\Lambda^{2n}}\left(1+O\left(\frac{r_D}r\right)\right)\\
&=\frac{r_D^{2(D-3)}}{r^{2D-2}}\Psi(D,r)\left(1+O\left(\frac{r_D}r\right)\right)
\end{split}
\label{EOMlambda111}
\end{equation}
Here again, we use an approximation that we are far away from the horizon and $\Psi(D,r)$ is a short-hand for the sum in the RHS of the equation.
Note that the zero term in the sum on the RHS of equation (\ref{EOMlambda111}) does not have $O(r_D/r)$ corrections.
Explicit form of coefficients $d_n$ is presented in Appendix~\ref{appA}, equations (\ref{dnD}), (\ref{alphaD}).
Here we write down the most essential coefficients for clarity. Namely
\begin{equation}
d_{n\geq2}(D)=\alpha(D)4^{n}\Gamma(n+\frac{D-1}2) \Gamma(n+D-1)=\frac{c_n(D)}{2n+D+1}
\end{equation}
and also note that $d_0(D)=c_0(D)$ and therefore $d_0(4)=0$ and $d_0(D>4)<0$ as well as $d_1(4)=126$.

Working out the series $\Psi(D,r)$ in the same way as $\Phi(D,r)$, namely splitting $n\geq2$ terms into a series $\Psi_2(D,r)$ and repeating the next steps in a similar fashion, one yields
\begin{equation}
    F'_{D}\left( r\right) =-\dfrac{P'_{D}\left( r\right) }{r^{D-1}}+\dfrac{2}{D-2}\dfrac{1}{M_{P}^{2}}\dfrac{r_{D}^{2\left( D-3\right) }}{r^{2D-1}}\left( d_{0}\left( D\right) {f_C}_{0}+\dfrac{d_{1}\left( D\right) {f_C}_{1}}{r^{2}\Lambda ^{2}}+\Psi _{2}\left(D,r\right) \right)
    \label{Fprime}
\end{equation}
where $\Psi_2(D,r)$ has the following integral representation through $\Fc_C(\tau)$
\begin{equation}
  \Psi_{2}\left( D,r\right)=2\alpha \left( D \right) \left(\frac{r\Lambda}2\right)^{\frac{3 D-3}{2}}\int ^{\infty }_{0}d\tau\left(\Fc_C(\tau)-{f_C}_0-{f_C}_1 \tau\right)\tau^{\frac{3D -7}{4}}K_{\frac{D-1}{2}}\left( r\Lambda\sqrt{\tau}\right)
  \label{psi2}
\end{equation}
Given that both $P'_D(r)$ and $P_D(r)$ have just been found explicitly above, finding $F_D(r)$ is a straightforward task. Using equation (\ref{Pprime}) to simplify (\ref{Fprime}), we see that prefactors in front of parentheses exactly match, and one can write
\begin{equation}
    F'_{D}\left( r\right) =\dfrac{2}{D-2}\dfrac{1}{M_{P}^{2}}\dfrac{r_{D}^{2\left( D-3\right) }}{r^{2D-1}}\left( \dfrac{(d_{1}\left( D\right)-c_{1}\left( D\right)) {f_C}_{1}}{r^{2}\Lambda ^{2}}+\Psi _{2}\left(D,r\right)- \Phi _{2}\left(D,r\right)\right)
    \label{Fprime1}
\end{equation}
where we have used the property of coefficients that $c_0(D)=d_0(D)$.
To get $F_D(r)$, we have to perform just one integration over $r$. Before an actual integration, however, it is instructive to note the following relation
\begin{equation*}
\begin{split}
    \Delta_2(D,r)=\Psi _{2}\left(D,r\right)- \Phi _{2}\left(D,r\right)&=
    2\alpha \left( D \right) \left(\frac{r\Lambda}2\right)^{\frac{3 D-3}{2}}\int ^{\infty }_{0}d\tau\left(\Fc_C(\tau)-{f_C}_0-{f_C}_1 \tau\right)\tau^{\frac{3D -7}{4}}\times\\
    &\times\left[K_{\frac{D-1}{2}}\left( r\Lambda\sqrt{\tau}\right)-r\Lambda\sqrt{\tau}K_{\frac{D-3}{2}}\left( r\Lambda\sqrt{\tau}\right)\right]
\end{split}
\end{equation*}
Finally one gets for $F_D(r)$
\begin{equation}
    F_{D}\left( r\right) =-\dfrac{2}{D-2}\dfrac{1}{M_{P}^{2}}\dfrac{r_{D}^{2\left( D-3\right) }}{r^{2D-1}}\left( \dfrac{(d_{1}\left( D\right)-c_{1}\left( D\right)) {f_C}_{1}}{r\Lambda ^{2}2D}+\tilde\Delta_2(D,r)\right)
    \label{Fans}
\end{equation}
where $\tilde\Delta_2(D, r)$ comes as the result of an integration of $\Delta_2(D, r)/r^{2D-1}$ with respect to $r$ with subsequent factoring of $-1/r^{2D-1}$ out. The corresponding general expression is fairly long and involves a generalized hypergeometric function ${}_1F_2$, and a generalized Meijer $G$-function, or logarithms. This problem arises because, upon accounting for all powers of $r$, the expression to be evaluated is
\begin{equation*}
\begin{split}
    \tilde\Delta_2(D,r)=&-2\alpha(D){2^{\frac{D+1}2}}\left(\frac{r\Lambda}{2}\right)^{2D-1}\int \frac{d\tau}\tau\bigg[\left(\Fc_C(\tau)-{f_C}_0-{f_C}_1 \tau\right){(\sqrt{\tau})^{2D-1}}\times\\
    &\qquad\qquad\qquad\qquad\qquad\qquad\qquad\quad\times\int x^{\nu}(K_{\nu+1}(x)-xK_{\nu+2}(x))dx\bigg]
    \\
    &\text{ with }\nu=-\frac D2-\frac12\text{ and }x=r\Lambda\sqrt{\tau}
    \end{split}
\end{equation*}
where we have to perform an integration over $x$, which turns out not to be as simple as in the above situation of finding $P_D(r)$. While we list the relevant integral (see equation (\ref{integral})), it seems to be more instructive to present an expression in $D=4$ which is as follows
\begin{equation}
\begin{split}
    \tilde\Delta_2(4,r)=&-\frac1{24}\left({r\Lambda}\right)^{7}\int {d\tau}\bigg[\left(\Fc_C(\tau)-{f_C}_0-{f_C}_1 \tau\right){{\tau}^{5/2}}\times\\
    &\qquad\qquad\qquad\qquad\times\left(4\text{Ei}(-x)+ e^{-x} \frac{4 x^2-x-1}{ x^3}\right)\bigg]
    \\
    &\text{ with }x=r\Lambda\sqrt{\tau}.
    \end{split}
\label{deltatilde2intD4}
\end{equation}
An integration constant must vanish in order not to modify the Newtonian potential at infinity.

To conclude this Section, we underline that we have constructed explicitly and analytically corrections to the Schwarzschild metric in first order in the coupling $\lambda$ far away from the horizon in arbitrary dimension $D$ for a selected set of EFT terms. Such terms are motivated by the study of asymptotically free gravity and by analyzing the effect of loop corrections. These terms represent operators that result in the most possible derivatives acting on a metric in equations of motion. This translates into taking into account the strongest corrections to the asymptotic behavior of the metric as long as the operators contain powers of the d'Alembertian. In the next Section we will see that a logarithm may provide an even stronger correction.

The notable fact is that thanks to a nice matching of computed coefficients by closed-form series and a subsequent integral rewriting of the sums, we can express corrections through the higher derivative form-factor in the model Lagrangian. Computations can be manifestly extended to all orders in $r_D/r$. On the way to deducing the presented equations, all such terms were explicitly computed; they are all obviously inverse powers of $r$ and thus can be readily integrated. This will give sub-leading contributions to the obtained results, but the actual result will most likely lack any nice closed form with no explicit summations. Moreover, other operators of the same mass dimension will give similar contributions, and in order to be consistent, we have to take into account other operators as well.

To illustrate this, consider a typical example term ${f_C}_1C_{\mu\nu\alpha\beta}(\Box^2/\Lambda^4) C_{\mu\nu\alpha\beta}$ in the action in four dimensions, which results in the following $\Delta^0_0$ component in the order $\lambda$
$$
\Delta^0_0=18{f_C}_4\frac{r_4^2}{r^{10}}\left(560  - 1488\frac{r_4}{r} + 939\frac{r_4^2}{r^2}\right)
$$
Similar computation for operators with higher powers of the d'Alembertian results in more terms of different degrees in parentheses.
Our analysis considers only the first term in the parentheses, the constant $560$. It is the most significant correction in the asymptotic infinity, and it can originate only upon accounting for the suggested higher derivative terms in the action. Other operators of the same mass dimension, for example $C_{\mu\nu\alpha\beta}C^{\mu\nu\sigma\rho}C_{\sigma\rho\gamma\delta}C^{\gamma\delta\alpha\beta}$ will result in the following correction
$$C_{\mu\nu\alpha\beta}C^{\mu\nu\sigma\rho}C_{\sigma\rho\gamma\delta}C^{\gamma\delta\alpha\beta}\overset{\sim}{\to}
\frac{r_4^3}{r^{11}} \left(64-67 \frac{r_4}r \right)
$$
It does not have the $\sim 1/r^{10}$ term but will impact the $\sim 1/r^{11}$ and higher order corrections, and thus would have to be added to the model for consistency if sub-leading asymptotics are to be computed.
We, however, do not expect any new effects upon accounting for these corrections. This situation is the same for all the EFT operators of the mass dimension $8$ or higher. One however would have similar contributions from all the operators of the mass dimension $6$ in four dimensions, i.e. $C_{\mu\nu\alpha\beta}\Box C^{\mu\nu\alpha\beta}$, $C_{\mu\nu\alpha\beta}C^{\mu\nu\sigma\rho}C_{\sigma\rho}^{\phantom{\sigma\rho}\alpha\beta}$ and others, which is in line with consideration of amplitudes in an EFT of gravity \cite{Ruhdorfer:2019qmk,deRham:2022gfe}.

On the technical side, we note that square in curvature operators without d'Alembertians do not contribute in 4 dimensions due to the topological nature of the Gauss-Bonnet term. Moreover, in any dimension, these operators give no contribution to the function $F_D(r)$. All other operators however do change $F_D(r)$ non-trivially making manifestly metric components $g_{00}$ and $g_{11}^{-1}$ different.

The next question to be explored in the subsequent analysis is the convergence of our results for metric corrections for different sets of ${f_C}_n$ and compatibility of form-factors with the perturbation scheme. Moreover, we will generalize our analysis to form-factors which do not have a Taylor expansion at zero. In particular, the logarithmic form-factor characteristic to 1-loop corrections will be studied by utilizing an integral representation for the logarithm.

\section{Resummation of the series and convergence analysis}


Here we aim at studying the convergence of the series $\Phi_2$ in (\ref{EOMlambda100}) emerging as the perturbative correction to the Schwarzschild solution. The series can be written as
\begin{equation}
    \Phi_{2}\left( \zeta\right) =\sum ^{\infty }_{n=2} c_n(D){f_C}_n\, \zeta^{n}.
\end{equation}
where we have defined
$\zeta=\frac{1}{r^2\Lambda^2}$.
Given that the coefficients $c_n$ contain Gamma-functions in the numerator, the series clearly has worse convergence properties compared to the expansion for form-factor (\ref{taylorF})
\begin{equation}
\Fc_C(\tau)=\sum_{n=0}^{\infty}{f_C}_n\,\tau^n.
\end{equation}
According to the Cauchy–Hadamard theorem, the power series is convergent for all values of $|\zeta|<\zeta_0$ with the convergence radius $\zeta_0$ given by
\begin{equation}
\label{convergence radius}
    \zeta_0 = \lim_{n \to \infty} |c_n(D){f_C}_n|^{\frac{1}{n}}.
\end{equation}

The behaviour of the coefficients ${f_C}_n$ for large $n$ is related to the asymptotic limit of the function $\Fc_C(\tau)$ at $\tau\to \infty$ in the complex plane of $\tau$. Any entire function $\phi(z)$ near the complex infinity can be characterized by an order
\begin{equation}
\label{order}
 \rho = \max \left( \lim_{|z| \to \infty} \frac{\log(\log |\phi(\zeta)|)}{\log |\zeta|}\right)= \lim_{n \to \infty} \frac{n \log n}{-\log a_n}, 
\end{equation}
and a type
\begin{equation}
\label{type}
    s = \max \left( \lim_{|z| \to \infty} \frac{\log |\phi(\zeta)|}{|\zeta|^\rho}\right)= \lim_{n \to \infty} \frac{\log \left( \frac{n}{e\, \rho} \right)}{|a_n|^{1/n}}.
\end{equation}
Here $a_n$ are the series coefficients, such that the entire function $\phi(\zeta)$ can be presented as
\begin{equation}
 \phi(\zeta)=\sum_{n=0}^{\infty}a_n \zeta^n \rightarrow  e^{s \zeta^{\rho}}\text{ when } |\zeta|\rightarrow \infty.
\end{equation}
In this way, the large $n$ limit of the series coefficients is related to the high-energy limit of the form-factor $\Fc_C(\Box)$ on the complex plane of the energy. It is interesting to note that this property is related to the locality of the underlying theory in the UV limit, and, in particular, to the polynomial boundedness of the scattering amplitudes. Indeed, substituting our result 
\begin{equation}
    c_{n\geq2}(D)=2\alpha(D)4^{n}\Gamma(n+\frac{D+1}2) \Gamma(n+D-1)
 \end{equation}
in equation \eqref{convergence radius} we find after expanding Gamma-functions for large $n$
\begin{equation}
    \zeta_0\sim \left(2 \pi  n^{3(D-1)/2} e^{2 n ( \log (2 n)-1)}\,{f_C}_n\right)^{-1/n}\sim  \frac{1}{n^2} {f_C}_n^{-1/n}.
\end{equation}
Here we take only the leading multiplier at $n\rightarrow\infty$. It is apparent that we have a non-zero convergence radius for the series only if 
\begin{equation}
    {f_C}_n^{-1/n}\gtrsim n^2,\quad {f_C}_n\lesssim n^{-2n}\sim\frac{1}{(n!)^2}.
\end{equation}
It means in particular that for an exponential type form-factor $\Fc(\Box)=e^{s\Box}$ with coefficients ${f_C}_n=1/(n!)>1/(n!)^2$ the series of corrections to the Schwarzschild solution is divergent everywhere.

The maximal growth rate of a form-factor on the complex energy plane which still results in a convergent series for corrections to a BH solution is of the form $e^{s \Box^{1/2}}$ which corresponds to order $\rho=1/2$ as defined in equation \eqref{order} while the type $s$ as defined in equation \eqref{type} can be arbitrary. It is interesting to mention that such a form-factor would also contribute to the graviton scattering amplitudes in a way corresponding to the localizable field theories \cite{Tokuda:2019nqb,Buoninfante:2023dyd,Buoninfante:2024ibt}. Namely, in such theories, amplitudes are polynomially bounded on the first sheet of the complex plane of total energy at fixed momentum transfer. Thus, we have come to a non-trivial conclusion that only in localizable theories do the perturbative corrections to the Schwarzschild metric make sense. For non-local exponential form-factors, the perturbative corrections form a series that is divergent everywhere except for the spatial infinity. 

\section{1-loop logarithm}
\label{seclog}

In this Section, we generalize our results to the case of form factors non-analytic at zero appearing in the quantum effective action as loop corrections. In particular, in $D=4$ we expect to have an operator
\begin{equation}
\Fc(\Box)=\gamma(\mu)\log{\frac{\Box}{\mu^2}}
\end{equation}
Here $\mu$ can be interpreted as a renormalization scale, and $\gamma(\mu)$ is a coupling defined at this scale.\footnote{Study of leading corrections in this case in the near-horizon limit was performed in \cite{Calmet:2025ocq}.}

We can understand the effect caused by this operator by proposing the analytic continuation of our result obtained for operators of the form $\Box^n$. To this end, we can write
\begin{equation}
    \log{\frac{\Box}{\mu^2}}=\lim_{n\rightarrow 0}\frac{(\Box/\mu^2)^n-1}{n}.
\end{equation}
This expression can be used as a definition of a non-local operator emerging at the loop level in the quantum effective action (see also \cite{Shaposhnikov:2022zhj,Shaposhnikov:2022dou} for more detailed discussion and representation of such an operator in the coordinate space). We use our result obtained for an infinite set of $\Box^n$ operators at integer $n\geq 2$,
\begin{equation}
\label{boxn}
  \Phi(D,r) =  2\alpha \left( D \right) \Gamma \left( n+\dfrac{D +1}{2}\right) \Gamma \left( n+D -1\right) \left(\frac{r^2\mu^2}{4}\right)^{-n},\quad \Fc(\Box)=(\Box/\mu^2)^n.
\end{equation}
Then we can define its analytic continuation in a variable $n$ to $n=0+\epsilon$, $\epsilon>0$. Here, we use such a procedure as a definition of a non-local logarithmic operator, leaving its more rigorous justification and relation to the coordinate space representation of effective action for future study.

Using equation \eqref{boxn} for the $\Box^n$ operator, we can take the formal limit $n\rightarrow 0$ (as it is prescribed by the analytic continuation) and find 
\begin{equation}
    \Phi_2(r)=\lim_{n\rightarrow 0}\gamma(\mu)\left(\frac{4 \Gamma \left(n+\frac{5}{2}\right) \Gamma (n+3) 4^n
   \left(\mu ^2 r^2\right)^{-n}}{n\,\sqrt{\pi }}-\frac{6}{n}\right)=- \gamma(\mu)\left( 12 \log{\mu r}+12\gamma_E-25\right).
\end{equation}
Then, from equation \eqref{Pprime} we get
\begin{equation}
   P'_4(r)= -\frac{r_4^2}{M_P^2 r^4} \gamma(\mu) \left(12 \log{\mu r}+12\gamma_E-25\right).
\end{equation}
Performing an integration, we obtain
\begin{equation}
\label{P4log}
   P_4(r)= \gamma(\mu)\frac{ r_4^2 \left( 4 \log \left(\mu  r\right)+4\gamma_E-7\right)}{ r^3
   M_P^2}.
\end{equation}
Remarkably, the same result can be obtained from equations \eqref{Panswer} and \eqref{phitilde2int} if one puts ${f_C}_0=0$, ${f_C}_1=0$ (although these series coefficients are not defined for the logarithm). Namely, the expression
\begin{equation}
    P_4(r)=-\gamma(\mu)\int_0^{\infty} d\tau \left( \frac{\mu^3 r_4^2 \tau^{5/4} (\mu r)^{3/2} \log (\tau)}{4 \sqrt{2 \pi }
   M_P^2}K_{3/2}(\mu r\sqrt{\tau})\right)=\gamma(\mu)\frac{ r_4^2 \left( 4 \log \left(\mu  r\right)+4\gamma_E-7\right)}{ r^3
   M_P^2}
\end{equation}
provides the same result as Eq. \eqref{P4log}. This is consistent with our definition of analytic continuation in $n$, and works for integer $n\geq 2$.

In a similar way, we can find the correction $F_4(r)$ from equation \eqref{EOMlambda111},
\begin{equation}
    M_P^2\left(\frac{F_4'(r)}{r}+\frac{P_4'(r)}{r^2}\right)=\gamma(\mu)\,\frac{r_4^2}{r^6}\,(-4 \log (\mu  r)-4 \gamma_E +7).
\end{equation}
Expressing $F_4'(r)$, one can obtain
\begin{equation}
    F_4'(r)=\frac{2\, \gamma(\mu)\,r_4^2}{M_P^2 r^5}(4\log{(\mu r)}+4\gamma_E-9).
\end{equation}
After integration, the correction to the metric finally takes the form
\begin{equation}
  F_4(r)=  -\frac{2\, \gamma(\mu)\,r_4^2}{M_P^2 r^4}(\log{(\mu r)}+\gamma_E-2).
\end{equation}
The terms omitted in this computation are suppressed by the extra powers of $r_4/r$, which makes them subdominant far away from the BH. It is interesting to notice here that the corrections in the metric induced by higher derivative EFT corrections (like $C_{\mu\nu\alpha\beta}\Fc(\Box/\Lambda^2) C^{\mu\nu\alpha\beta}$) are always more suppressed than the ones from $C_{\mu\nu\alpha\beta}\log({\Box}/\mu^2)C^{\mu\nu\alpha\beta}$ operator corresponding to the loop effects in GR. Stating this, we assume that the form-factor $\Fc(\tau)$ is compatible with the resummation of corrections as discussed in the previous Section. This situation is specific for four dimensions because the effects of the tree-level quadratic in curvature terms are vanishing due to the topological nature of the Gauss-Bonnet term. This means that logarithmic loop corrections can result in a more significant deformation of the Schwarzschild solution in four dimensions.

\section{Conclusions}

In this work, we aimed at computing a deformation of the Schwarzschild solution in the presence of a sequence of higher derivative EFT operators up to an infinite order in derivatives. We particularly focus on operators of the form  $C_{\mu\nu\alpha\beta} (\Box^n/\Lambda^{2n}) C^{\mu\nu\alpha\beta}$ and study the corresponding deformations of the vacuum BH solution of GR perturbatively in the EFT coefficients. For each power $n$, we find an explicit expression for corrections to the metric in an arbitrary space-time dimension $D\geq 4$. Moreover, for $n\geq 2$, based on the xAct (\href{https://xact.es/index.html}{\texttt{https://xact.es/index.html}}) computation done for $n\leq 10$, we observe a regular pattern of coefficients in the leading (most slowly decaying) order in $r_D/r$ expansion. This allows us to write a general formula for these corrections. This result further leads to an integral representation for the corrections in the presence of a generic operator $C_{\mu\nu\alpha\beta} \Fc(\Box/\Lambda^2) C^{\mu\nu\alpha\beta}$ solely through the function $\Fc(\tau)$. We provide analytic expressions for the corrected metric components under the assumption that the function $\Fc(\tau)$ can be expanded in a Taylor series around zero. However, we find that the deformation expressed in power series is not always convergent, which puts restrictions on the use of the perturbation theory methods for searching for a solution with an infinite tower of higher-derivative EFT operators. Our main result is the observation that the very possibility of finding corrections to a BH solution in the form of a perturbative deformation of the Schwarzschild solution is related to UV locality of the underlying theory. Namely, the term $C_{\mu\nu\alpha\beta}\Fc(\Box/\Lambda^2) C^{\mu\nu\alpha\beta}$ would contribute to the scattering amplitude of four gravitons, and its UV asymptotic. In non-localizable theories, the maximal growth of the amplitudes on the complex energy plane is larger than $e^{\alpha|p|}$. The corresponding form-factors $\Fc(\tau)$ would cause a deformation of the metric in the form of series which are divergent everywhere except for the spatial infinity $r\rightarrow\infty$.

Even though our results are obtained under the assumption that the function $\Fc(\tau)$ can be expanded in a Taylor series around zero, we can also perform an analytic continuation with respect to the power $n$ of the $\Box^n$ operator. Having a closed-form expression of corrections to the metric from $C_{\mu\nu\alpha\beta}(\Box^n/\Lambda^{2n}) C^{\mu\nu\alpha\beta}$ operator as a function of an integer power $n\geq 2$, we can use it for fractional powers and the logarithm in the form-factor $\Fc(\tau)$. Such an extension allows for computing corrections to the vacuum solution in GR from graviton loops, summarised in a 1-loop effective action as $C_{\mu\nu\alpha\beta}\log{(\Box/\mu^2)}C^{\mu\nu\alpha\beta}$ operator in four dimensions. We compute an explicit deformation of the Schwarzchild solution for such an operator and have found that it is dominant at large $r\gg r_4$ compared to the one obtained from the EFT corrections as long as EFT coefficients lead to a convergent resummation of corrections. This situation is specific to four dimensions because there are no local EFT terms quadratic in curvature, which would correct the solution.

Our techniques and results can be straightforwardly used for computing corrections to BH solutions in asymptotically safe quantum gravity. In this context, the leading deformation far away from the BH comes from the part of the action determining the non-perturbative graviton propagator. Our result provides a technique which can be used for such a computation, given the explicit form of the graviton two-point function is known.
Looking forward, the presented technique can be extended to other vacuum solutions of GR, and the first future step will be to expand it to the case of a rotating Kerr BH metric. Other configurations, like a charged Reissner-Nordstrom BH, can also be considered in this framework. One of the most interesting limiting cases would be extremal configurations and constraints coming from the weak gravity conjecture \cite{Arkani-Hamed:2006emk,Arkani-Hamed:2021ajd}.

\subsection*{Acknowledgements}

The work of A.~T. was supported by the National Natural Science Foundation of China (NSFC) under Grant No. 12547104 and No. 12505091. 

\appendix

\section{Explicit form of equation (\ref{EOMlambda1})}
\label{appA}

\subsection{General form of $\left.G^\mu_\nu\right|_{ds_\lambda^2}$}

\begin{align}
G^0_0|_{ds^2_\lambda}&=\lambda\frac{D-2}2\frac{{P_D}'(r)}{r^{D-2}},\\
G^1_1|_{ds^2_\lambda}&=\lambda\frac{D-2}2\frac{(r^{D-3}-{r_D}^{D-3}) {F_D}'(r)}{r^{D-2}}+G^0_0|_{ds^2_\lambda}+O\left(\lambda ^2\right),\\
G^i_i|_{ds^2_\lambda}&=\frac\lambda2\left[\frac{(r^{D-3}-{r_D}^{D-3}) {F_D}''(r)+{P_D}''(r)}{r^{D-3}}+\frac{D-3}2\frac{(2 r^{D-3}+{r_{D}}^{D-3}) {F_D}'(r)}{r^{D-2}}\right]+O\left(\lambda ^2\right)
\end{align}
Here index $i$ runs from 2 to $D-1$, and there is no sum over $i$.

\subsection{Coefficients $c_n(D)$ and $d_n(D)$}

\noindent\textbf{Case $D=4$}

$c_0=0$, $c_{n}$ starting from $c_0$ are:
\begin{equation*}
\begin{split}
0,90,&10080,907200,119750400,21794572800,5230697472000,1600593426432000,\\
&608225502044160000,281000181944401920000,155112100433309859840000,\dots  
\end{split}
\end{equation*}
Note that $c_0=0$ holds thanks to the fact that the Gauss-Bonnet term in $D=4$ is topological and thus does not contribute to equations as long as the boundary term is not accounted for.
Using an analytic matching starting from $c_2$, the coefficients can be expressed as
\begin{equation}
    c_0=0,~c_1=90,~c_{n>=2}=\frac1{\sqrt{\pi}}4^{n+1}\Gamma(n+\frac52)\Gamma(n+3)=\frac1{4} \Gamma(2n+5)
\end{equation}

$d_0=0$, $d_{n}$ starting from $d_0$ are:
\begin{equation*}
\begin{split}
0,126,&1440,100800,10886400,1676505600,348713164800,94152554496000,\\
&32011868528640000,13380961044971520000,6744004366665646080000\dots
\end{split}
\end{equation*}
Using an analytic matching starting from $d_2$, the coefficients can be expressed as
\begin{equation}
d_0=0,~d_1=126,~d_{n\geq2}=2\frac1{\sqrt{\pi}}\cdot 4^n \Gamma(n + \frac32) \Gamma(n + 3)
\end{equation}

\noindent\textbf{Case $D=5$}

$c_0=-12$, $c_{n}$ starting from $c_0$ are:
\begin{equation*}
\begin{split}
 -12,
 1440,&
 184320,
 22118400,
 3715891200,
 832359628800,
 239719573094400,
 86299046313984000,\\
 &37971580378152960000,
 20048994439664762880000,
 12510572530350812037120000,\dots
\end{split}
\end{equation*}
Using an analytic matching starting from $c_2$, the coefficients can be expressed as
\begin{equation}
    c_0=-12,~c_1=1440,~c_{n>=2}=4^{n+1} \Gamma(n+3) \Gamma(n+4)
\end{equation}

$d_0=-12$, $d_{n}$ starting from $d_0$ are:
\begin{equation*}
\begin{split}
-12,1248,&23040,2211840,309657600,59454259200,14982473318400\\
&4794391461888000,1898579018907648000,\\
&911317929075671040000,521273855431283834880000,\dots
\end{split}
\end{equation*}
Using an analytic matching starting from $d_2$, the coefficients can be expressed as
\begin{equation}
d_0=-12,~d_1=1248,~d_{n\geq2}=2\cdot 4^n \Gamma(n + 2) \Gamma(n + 4)
\end{equation}

\noindent\textbf{Case $D=6$}

$c_0=-60$, $c_{n}$ starting from $c_0$ are:
\begin{equation*}
\begin{split}
 -60,
 8400,&
 1360800,
 209563200,
 43589145600,
 11769069312000,
 4001483566080000,\\
 &1672620130621440000,
 843000545833205760000,\\
 &504114326408257044480000,
352880028485779931136000000,
\end{split}
\end{equation*}
Using an analytic matching starting from $c_2$, the coefficients can be expressed as
\begin{equation}
c_0=-60,~c_1=8400,~c_{n\geq2}=\frac1{\sqrt{\pi}}{4^{n+1}}\Gamma(n+\frac72)\Gamma(n+5)
\end{equation}

$d_0=-60$, $d_{n}$ starting from $d_0$ are:
\begin{equation*}
\begin{split}
-60,6000,&151200,19051200,3353011200,784604620800,235381386240000,\\
&88032638453760000,40142883134914560000,\\
&21918014191663349760000,14115201139431197245440000,\dots
\end{split}
\end{equation*}
Using an analytic matching starting from $d_2$, the coefficients can be expressed as
\begin{equation}
d_0=-60,~d_1=6000,~d_{n\geq2}=2 \frac1{\sqrt{\pi}}4^n \Gamma(n + \frac52) \Gamma(n + 5)
\end{equation}

\noindent\textbf{Case $D=7$}

$c_0=-180$, $c_{n}$ starting from $c_0$ are:
\begin{equation*}
\begin{split}
-180,31680,&
6451200,1238630400,312134860800,99883155456000,39553729560576000\\
&18985790189076480000,10859871988151746560000\\
&7297833976037973688320000,5692310501309619476889600000\dots  
\end{split}
\end{equation*}
Using an analytic matching starting from $c_2$, the coefficients can be expressed as
\begin{equation}
    c_0=-180,~c_1=31680,~c_{n>=2}=\frac16 4^{n+1} \Gamma(n+4) \Gamma(n+6)
\end{equation}

$d_0=-180$, $d_{n}$ starting from $d_0$ are:
\begin{equation*}
\begin{split}
-180,20160,&645120, 103219200, 22295347200, 6242697216000, 2197429420032000,\\
&949289509453824000,493630544915988480000,\\
&304076415668248903680000,218935019281139210649600000,\dots
\end{split}
\end{equation*}
Using an analytic matching starting from $d_2$, the coefficients can be expressed as
\begin{equation}
d_0=-180,~d_1=20160,~d_{n\geq2}=\frac13 4^n \Gamma(n + 3) \Gamma(n + 6)
\end{equation}

\noindent\textbf{Case $D=8$}

$c_0=-420$, $c_{n}$ starting from $c_0$ are:
\begin{equation*}
\begin{split}
-420,92610,&23284800,5448643200,1634592960000,611337767040000,278770021770240000,\\
&152208431886551040000,\dots  
\end{split}
\end{equation*}
Using an analytic matching starting from $c_2$, the coefficients can be expressed as
\begin{equation}
    c_0=-420,~c_1=92610,~c_{n>=2}=\frac1{18}\frac1{\sqrt{\pi}} 4^{n+1} \Gamma(n+\frac92) \Gamma(n+7)
\end{equation}

$d_0=-420$, $d_{n}$ starting from $d_0$ are:
\begin{equation*}
\begin{split}
-420,54390,&2116800,419126400,108972864000,35961045120000,14672106408960000,\\
&7248020566026240000,\dots
\end{split}
\end{equation*}
Using an analytic matching starting from $d_2$, the coefficients can be expressed as
\begin{equation}
d_0=-420,~d_1=54390,~d_{n\geq2}=\frac19 \frac1{\sqrt{\pi}}4^n \Gamma(n + \frac72) \Gamma(n + 7)
\end{equation}

\noindent\textbf{Case $D=9$}

$c_0=-840$, $c_{n}$ starting from $c_0$ are:
\begin{equation*}
\begin{split}
-840,228480,&69672960, 19508428800, 6866966937600,\dots  
\end{split}
\end{equation*}
Using an analytic matching starting from $c_2$, the coefficients can be expressed as
\begin{equation}
    c_0=-840,~c_1=228480,~c_{n>=2}=\frac1{240} 4^{n+1} \Gamma(n+5) \Gamma(n+8)
\end{equation}

$d_0=-840$, $d_{n}$ starting from $d_0$ are:
\begin{equation*}
\begin{split}
 -840,
 126336,
 5806080,
1393459200,
429185433600,\dots
\end{split}
\end{equation*}
Using an analytic matching starting from $d_2$, the coefficients can be expressed as
\begin{equation}
d_0=-840,~d_1=126336,~d_{n\geq2}=\frac1{120} 4^n \Gamma(n + 4) \Gamma(n + 8)
\end{equation}

\noindent\textbf{Case $D=10$}

$c_0=-1512$, $c_{n}$ starting from $c_0$ are:
\begin{equation*}
\begin{split}
-1512,498960,&181621440,59935075200,24453510681600,\dots  
\end{split}
\end{equation*}
Using an analytic matching starting from $c_2$, the coefficients can be expressed as
\begin{equation}
    c_0=-1512,~c_1=498960,~c_{n>=2}=\frac1{1350}\frac1{\sqrt{\pi}} 4^{n+1} \Gamma(n+\frac{11}2) \Gamma(n+9)
\end{equation}

$d_0=-1512$, $d_{n}$ starting from $d_0$ are:
\begin{equation*}
\begin{split}
 -1512,263088,&13970880,3995671680,1438441804800,\dots
\dots
\end{split}
\end{equation*}
Using an analytic matching starting from $d_2$, the coefficients can be expressed as
\begin{equation}
d_0=-840,~d_1=263088,~d_{n\geq2}=\frac1{675} \frac1{\sqrt{\pi}}4^n \Gamma(n + \frac92) \Gamma(n + 9)
\end{equation}

\noindent\textbf{General pattern}

The patterns appearing and looking very plausible are as follows.

For $c_n$:
\begin{equation}
\label{cnD}
\begin{split}
    c_0(D)&=-\frac12(D-4)(D-3)(D-2)(D-1)\\
    c_1(D)&=90,1440,8400,31680,92610,228480,498960,\dots\\
    c_{n\geq2}(D)&=2\alpha(D)4^{n}\Gamma(n+\frac{D+1}2) \Gamma(n+D-1)
\end{split}
\end{equation}

For $d_n$:
\begin{equation}
\label{dnD}
\begin{split}
    d_0(D)&=c_0(D)=-\frac12(D-4)(D-3)(D-2)(D-1)\\
    d_1(D)&=126,1248,6000,20160,54390,126336,263088,\dots\\
    d_{n\geq2}(D)&=\alpha(D)4^{n}\Gamma(n+\frac{D-1}2) \Gamma(n+D-1)
\end{split}
\end{equation}
Here $\alpha(D)$ takes the values
\begin{equation}
\begin{split}
\label{alphaD}
    \alpha(D)&=\frac2{\sqrt{\pi}},\frac2{\sqrt{\pi}},\frac1{9\sqrt{\pi}}, \frac1{675\sqrt{\pi}}\text{ in }D=4,6,8,10\\
    \alpha(D)&=2,\frac13,\frac1{120} \text{ in }D=5,7,9
\end{split}
\end{equation}
and both $c_1(D)$ and $d_1(D)$ (as well as $c_1(D)-d_1(D)$) do not obey an easily recognizable pattern.
We also notice that for $n\geq2$ one has
$$c_n(D)=2\left(n+\frac{D+1}2\right)d_n(D).$$

\section{Integrals involving modified Bessel function $K_\nu(x)$}

Integrals listed below can be found in  \cite{Gradshteyn:1702455} or evaluated using Wolfram Mathematica.
An integral used to transform (\ref{totransform}) is as follows
\begin{equation}
\int_0^\infty x^\mu K_\nu(ax)dx=2^{\mu-1}a^{-\mu-1}\Gamma\left(\frac{1+\mu+\nu}2\right)\Gamma\left(\frac{1+\mu-\nu}2\right)
\label{intgg}
\end{equation}
A standard integral relevant for our analysis is
\begin{equation}
    \int x^\nu K_{\nu-1}(x)dx=-x^\nu K_{\nu}(x)
    \label{intsimple}
\end{equation}
Another integral relevant for our analysis is
\begin{equation}
\begin{split}
    \int x^\nu K_{\nu+1}(x)dx=
    \pi  2^{-\nu -3} \frac1{\sin (\pi  \nu )} &\left(\frac{4^{\nu }\left( 8\nu  \log (x)-x^2 \, _2F_3\left(1,1;2,2,1-\nu ;\frac{x^2}{4}\right)\right)}{\Gamma (1-\nu )}\right.\\
    &\left.+\frac{x^{2 \nu +2} \, _1F_2\left(\nu +1;\nu +2,\nu
   +2;\frac{x^2}{4}\right)}{(\nu +1)^2 \Gamma (\nu +1)}\right)
   \end{split}
   \label{integral}
\end{equation}
In this case, we have $\nu=-D/2-1/2$. Then, for even dimensions, one can find
\begin{equation}
\begin{split}
    \left.\int (x^\nu K_{\nu+1}(x)-x^{\nu+1} K_{\nu+2}(x))dx\right|_{D=4}=
\frac13\sqrt{\frac{\pi }{2}}\left(4\text{Ei}(-x)+ e^{-x} \frac{4 x^2-x-1}{ x^3}\right)
   \end{split}
\end{equation}
\begin{equation}
\begin{split}
    \left.\int (x^\nu K_{\nu+1}(x)-x^{\nu+1} K_{\nu+2}(x))dx\right|_{D=6}=\frac15\sqrt{\frac{\pi }{2}}\left(-2\text{Ei}(-x)-e^{-x}\frac{2 x^4-2 x^3-x^2+3 x+3}{x^5}\right)
   \end{split}
\end{equation}
and a similar pattern containing exponent, integral exponents, and fractions repeats for higher even $D$. For odd $D$, the situation is a bit more complicated as the first integral (\ref{integral}) is not very convenient given an apparent singularity for integer $\nu$ due to the sine in the denominator. However, an alternative form using Meijer $G$-function can be found. The following is the result for $D=5$
\begin{equation}
\begin{split}
    &\left.\int (x^\nu K_{\nu+1}(x)-x^{\nu+1} K_{\nu+2}(x))dx\right|_{D=5}\\
    =&
\frac{1}{16} \left(2 G_{1,3}^{3,0}\left(\left.\frac{x}{2},\frac{1}{2}\right|
\begin{array}{c}
 1 \\
 -1,0,0 \\
\end{array}
\right)-G_{1,3}^{3,0}\left(\left.\frac{x}{2},\frac{1}{2}\right|
\begin{array}{c}
 1 \\
 -2,0,0 \\
\end{array}
\right)\right)
\end{split}
\end{equation}
where the generalized Meijer $G$-function is defined as follows
\begin{equation*}
G^{m,n}_{p,q}\left(z,r\bigg|
\begin{array}{c}
 a_1,\dots,a_p \\
 b_1,\dots,b_q \\
\end{array}
\right)=
    \frac r{2\pi i}\int\frac{\prod_{i=1}^n \left(-a_i-r s+1\right) \prod_{j=1}^m\left(b_j+r s\right)}{\prod_{i=n+1}^p\left(a_i+r s\right)\prod_{j=1+m}^q
   \left(-b_j-r s+1\right)}z^{-s}ds
\end{equation*}



\providecommand{\href}[2]{#2}\begingroup\raggedright\endgroup
\end{document}